# The Source of Turbulence in Astrophysical Disks:

# An Ill-posed Problem.

Denis Richard,
NASA Ames Research Center

UCSC-Ames Planet and Star Formation Meeting
March 29, 2007

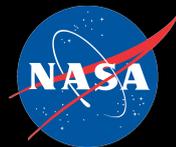

# Astrophysical Disks

**Disks are ubiquitous in Astrophysics :**

* Planetary disks

* Circumstellar Disks (around young stars)

* Binary systems

* Active Galactic Nuclei (around black holes)

**Therefore, understanding disks is fundamental to understand planetary and stellar formation and evolution.**

**Wide range of sizes :**

From Saturn's rings ( $\sim 10^7$ km) to AGN disks ($\sim$ parsec = $3.1 \, 10^{13}$ km)

**Wide variety of complex physical processes,**
one of them being the transport of angular momentum.

# Astrophysical Disks (Artist view)

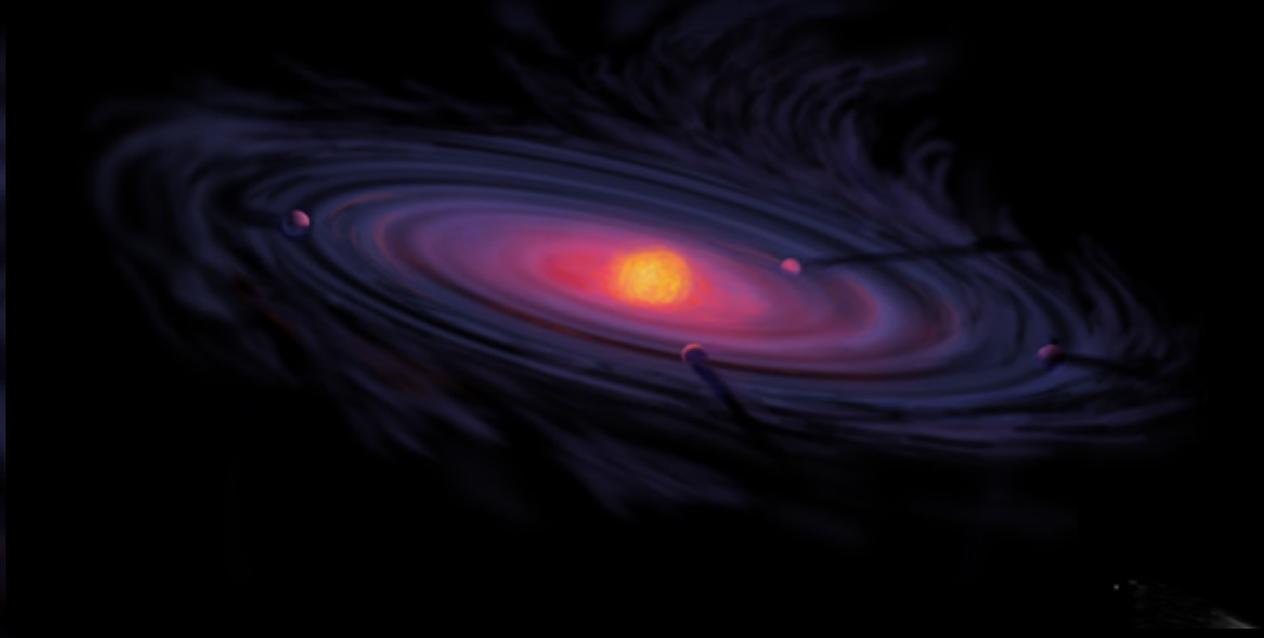
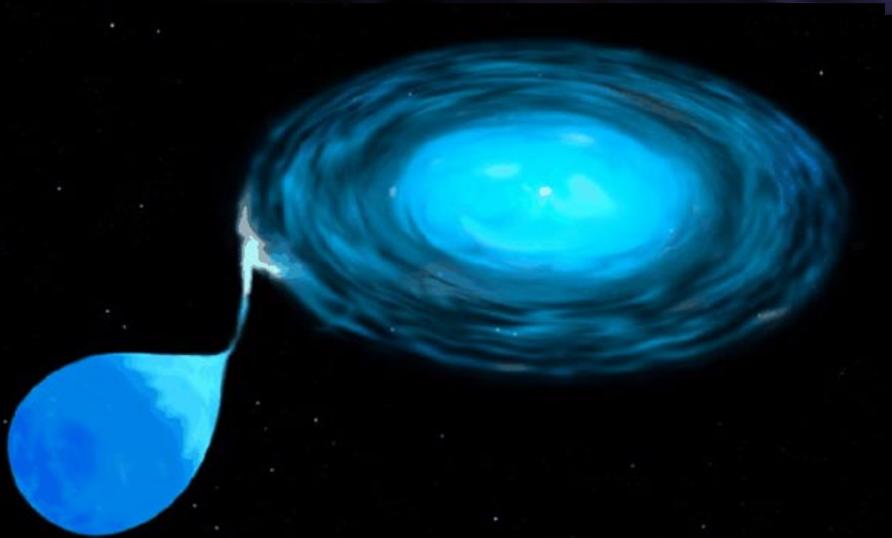
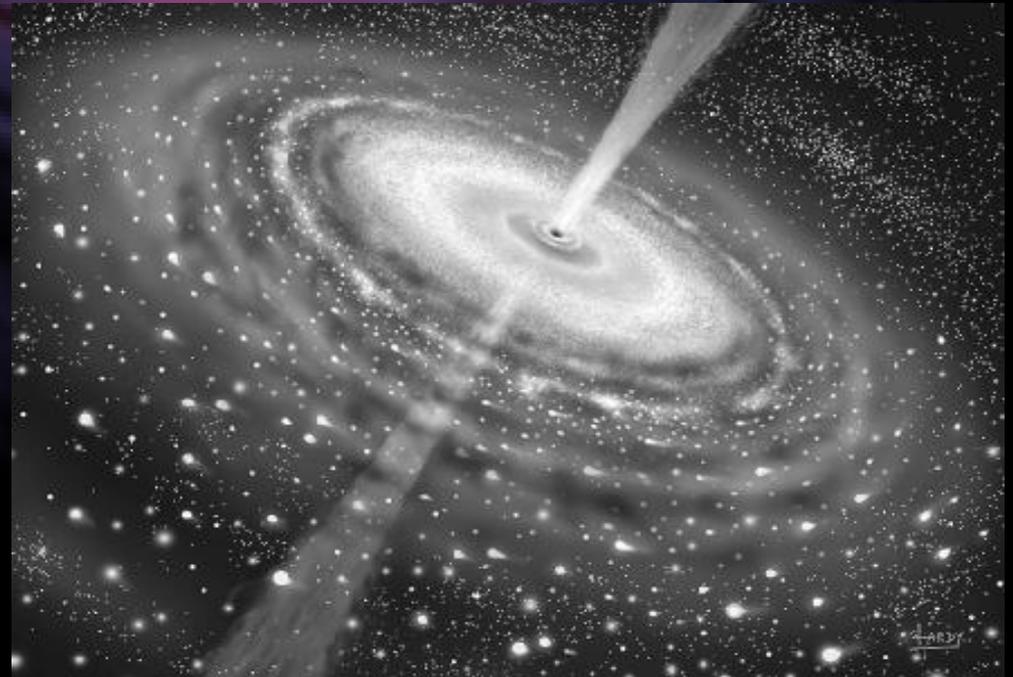

# Astrophysical Disks (HST)

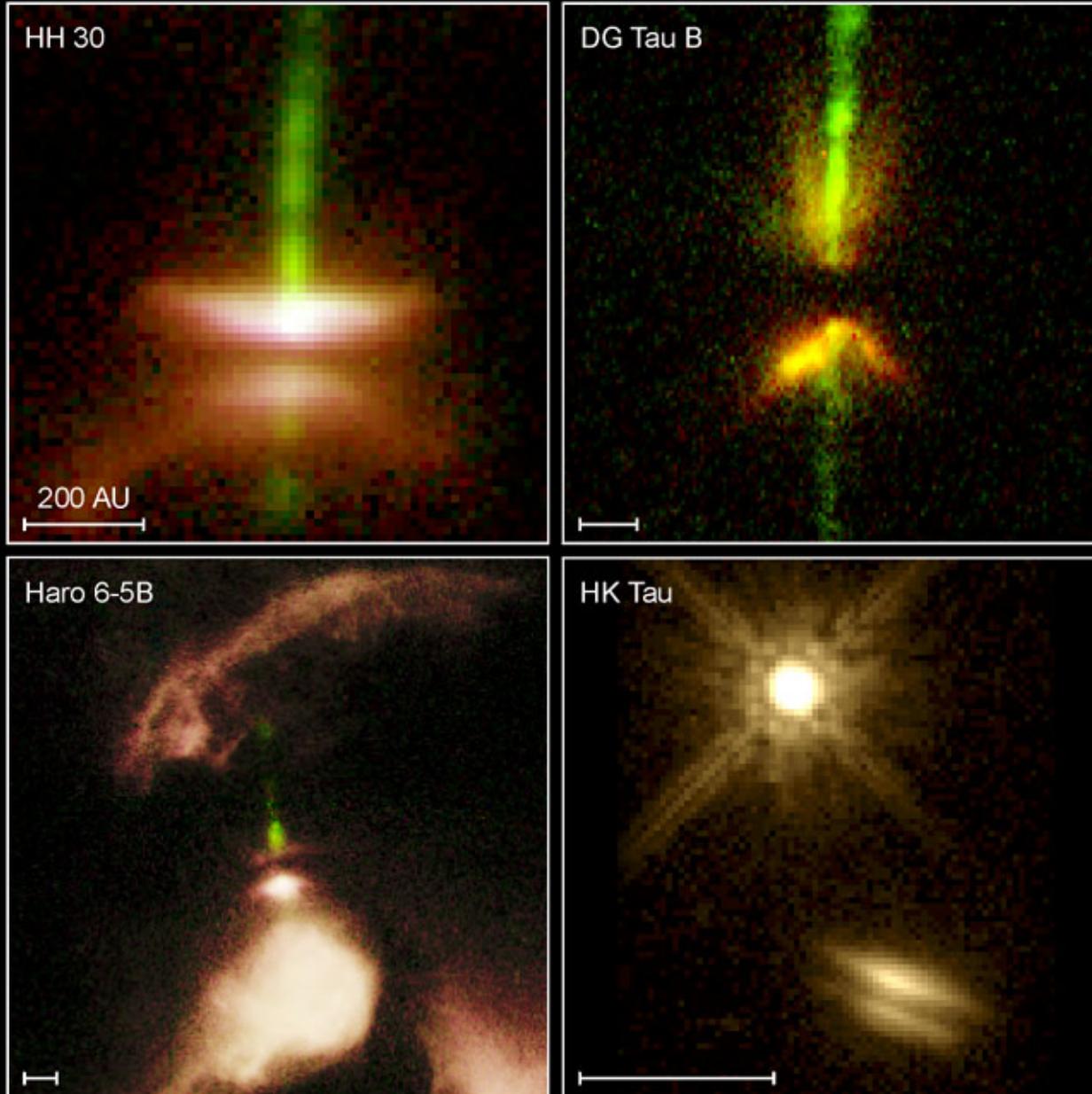

# Turbulence in Disks

* **Accretion** Disk : gas and "dust" falling inward, toward the central object.

   * **Thin disk** = H/R << 1   →   **Keplerian rotation** : $\Omega \sim r^{-3/2}$ (no radial or vertical velocity in first approximation)

   * To maintain stationary rotation, **angular momentum** needs to be **transported outward.**

* **Need for an adequate transport mechanism.**

   * Molecular viscosity is too small.

   *  Turbulence :  instability mechanism ? Transport properties ?

* **Early models** : **Shakura & Sunyaev 1973 :**

   * Turbulence most likely generated by **differential rotation** (shear flow).

   * *Ad hoc model* for transport : **turbulent viscosity**  based on smallest **constraints** :

$$v_t = \alpha \cdot Cs \cdot H$$

# A Short History of Turbulence Models for Accretion Disks

* **1973 – 1991 : Shakura & Sunyaev era :**

  - Source of turbulence unknown (shear flow assumption ?)

  - Work on turbulence, in order to improve transport model.

  - meanwhile, transport model : $v_t = \alpha \cdot Cs \cdot H$

* **1991 : Magneto Rotational Instability** rediscovered
  Weak magnetic field coupled to shear flow gives rise to a linear instability.
  (Chandrasekhar 1960, Balbus & Hawley 1991)

* **1991 – present : MRI era :**

  - Source of turbulence : MRI

  - Turbulent model : $v_t = \alpha \cdot Cs \cdot H$

# Turbulent Viscosity Model

**What is it ?**

   * **Analytically :** A description of turbulent transport as a *diffusive mechanism*.

   * For **numerical simulation** : a type of basic *subgrid model*.

**Where does it come from ?**

   * Built *ad hoc* (alpha-viscosity) : *relevant length scale* x *relevant velocity*.
   or
   * Measured experimentally (lab or numerical) : Reynold stress.

**When should it be used ?**

   * When studying anything **BUT** turbulence as a fundamental physical process.

   * When simulations do not have adequate resolution to describe the whole range of scales of the flow, in which case the viscosity model has to be chosen as to describe **only the subgrid scales.**

# From Instabilities to Turbulence

**Two types of Instabilities :**

* **Linear** : flow unstable to infinetesimal perturbations (super-critical transitions)
    ex : thermal convection, Rayleigh/centrifugal instability in rotating flows.

* **Non-Linear** : flow unstable to finite amplitudes (sub-critical transitions)
    ex : plan shear flow, Differential rotation (?)

**Analytical :** Linear can be well treated (transition) / no general model for Non-Linear.

**Numerical :** Linear : generally low Reynolds, large scale flows = "lower" resolution
    Non-linear : generally high Reynolds, small scale flows = "high" resolution

**Laboratory :** In theory can study both types equally well.

**In an Astrophysical context :** Both types are equally difficult to study,
    because what ultimately matters is the **turbulent state**, which will generally be at very
    high Reynolds, thus difficult to describe.

Thus, while **MRI** has been around for more than 15 years in the disk community, there is no associated description for turbulent transport. An instability easy to describe does not mean that the induced turbulence is equally easy to quantify.

# Polemic ? What Polemic ?

* Little to no doubt that **MRI is at work in most accretion disks.**

* But : *Is MRI the only turbulent mechanism relevant to Astrophysical disks ?*

<u>"Schools of thoughts"</u> :

**MRI school**

* MRI is necessary to power accretion disks
(Only MRI can provide adequate angular momentum transport.)

* MRI is sufficient to power all accretion disks
(giving birth to such things as "Dead zone models")

**Instability X school**

* Instability X is also relevant to Astrophysical Disks dynamics.
(Differential Rotation, Plane Shear, Strato-rotational, Baroclinic,...)

**No-school school**

* Just tell me how turbulence acts in my disk, so that I can react, coagulate, form a planet, evolve, etc...

# Arguments against Differential Rotation

**Analytical**: radial and azimuthal fluctuations can not grow at the same time.
(Balbus, Hawley & Stone, 1996)

$$\frac{\partial}{\partial t}\left\langle\frac{\rho u_R^2}{2}\right\rangle = 2\Omega\langle\rho u_R u_\phi\rangle - \left\langle u_R\frac{\partial P}{\partial R}\right\rangle - \langle\rho\nu|\nabla u_R|^2\rangle, \quad (2.6a)$$

$$\frac{\partial}{\partial t}\left\langle\frac{\rho u_\phi^2}{2}\right\rangle = -\frac{\langle\rho u_R u_\phi\rangle}{R}\frac{dR^2\Omega}{dR} - \left\langle\frac{u_\phi}{R}\frac{\partial P}{\partial \phi}\right\rangle - \langle\rho\nu|\nabla u_\phi|^2\rangle. \quad (2.6b)$$

Problem: this set of equations is linear, thus irrelevant to non-linear instabilities (~ Rayleigh criterion)

**Numerical**: No-show in numerical simulations
(Balbus, Hawley & Stone, 1996)

Problem: Reynolds numbers are "low" (a $1024^3$ grid can simulate a maximum Re of order 10,000)

**Experimental**: H.Ji, M.Burin, E.Schartman & J.Goodman, 2006 (accompanied by comments by S.Balbus)

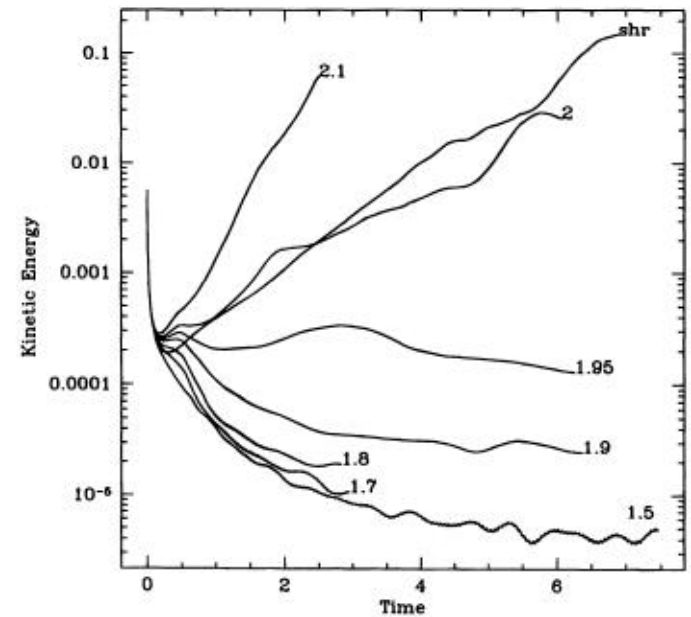

FIG. 1.—Evolution of kinetic energy fluctuations, in units of $(L\Omega)^2$, for different rotation profiles. Curves are labeled by the power-law index $q$ of the angular velocity $\Omega(r) \sim r^{-q}$. The curve labeled "shr" corresponds to pure Cartesian shear. The $q = 2.1$ curve is Rayleigh unstable, the $q = 2$ and "shr" curves are nonlinearly unstable, and the rest are stable. The Keplerian profile is $q = 1.5$. Note the dramatic difference in the evolution of the $q = 2$ and $q = 1.95$ profiles.

# Can Differential Rotation lead to Turbulence in Disks ?

* **High Reynolds Shear flow** : Astrophysical Disks Re ~ $10^6$ to $10^{20}$

* **Early (1930's) laboratory experiments** :
  Couette-Taylor flow unstable at high Re.
  (Richard & Zahn, 1999)

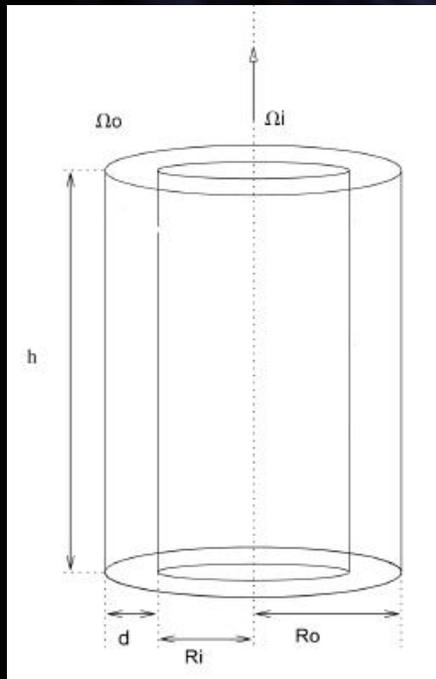

$$Re^* = \frac{R^3}{\nu}\frac{\Delta\Omega}{\Delta R} \geq Re_c^* \simeq 6\,10^5.$$

$$\nu_t = \beta r^3 \left|\frac{d\Omega}{dr}\right|,$$

$$\beta = 1.5 \pm 0.5\,10^{-5}.$$

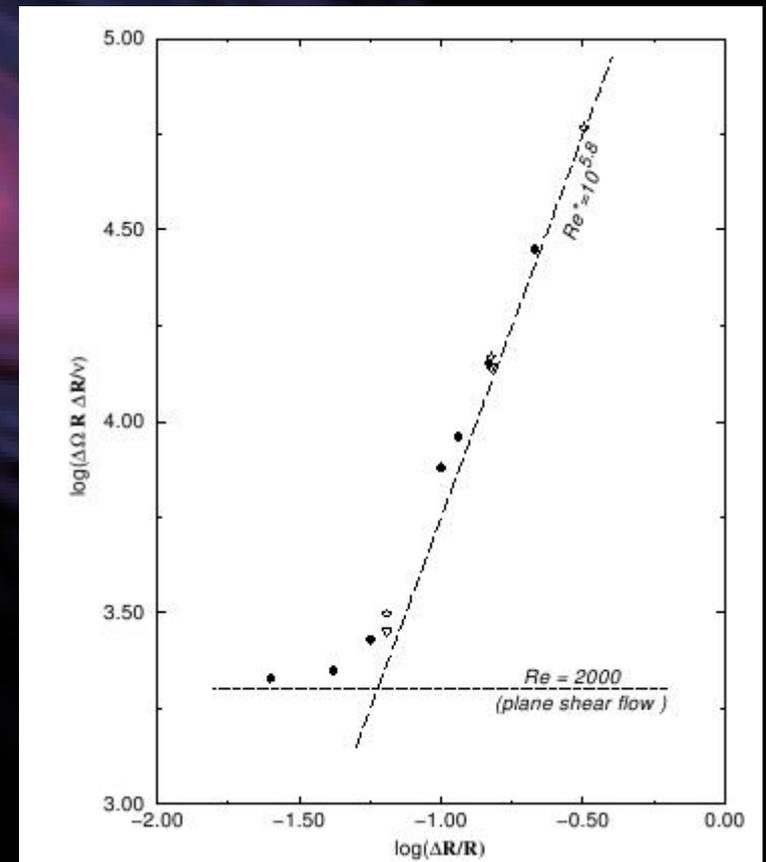

Data from Wendt (1933) and Taylor (1936)

# Can Differential Rotation lead to Turbulence in Disks ?

* **New experimental setup** (Richard, 2001).

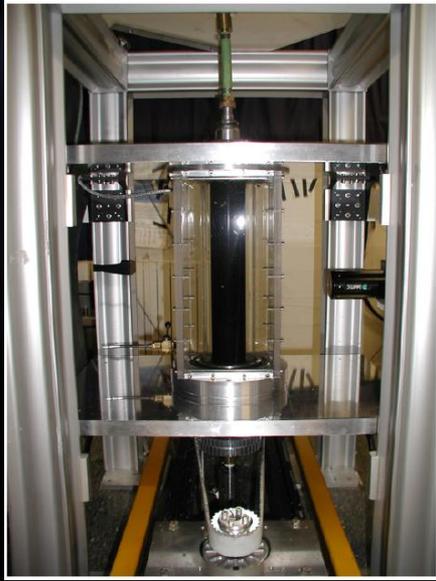

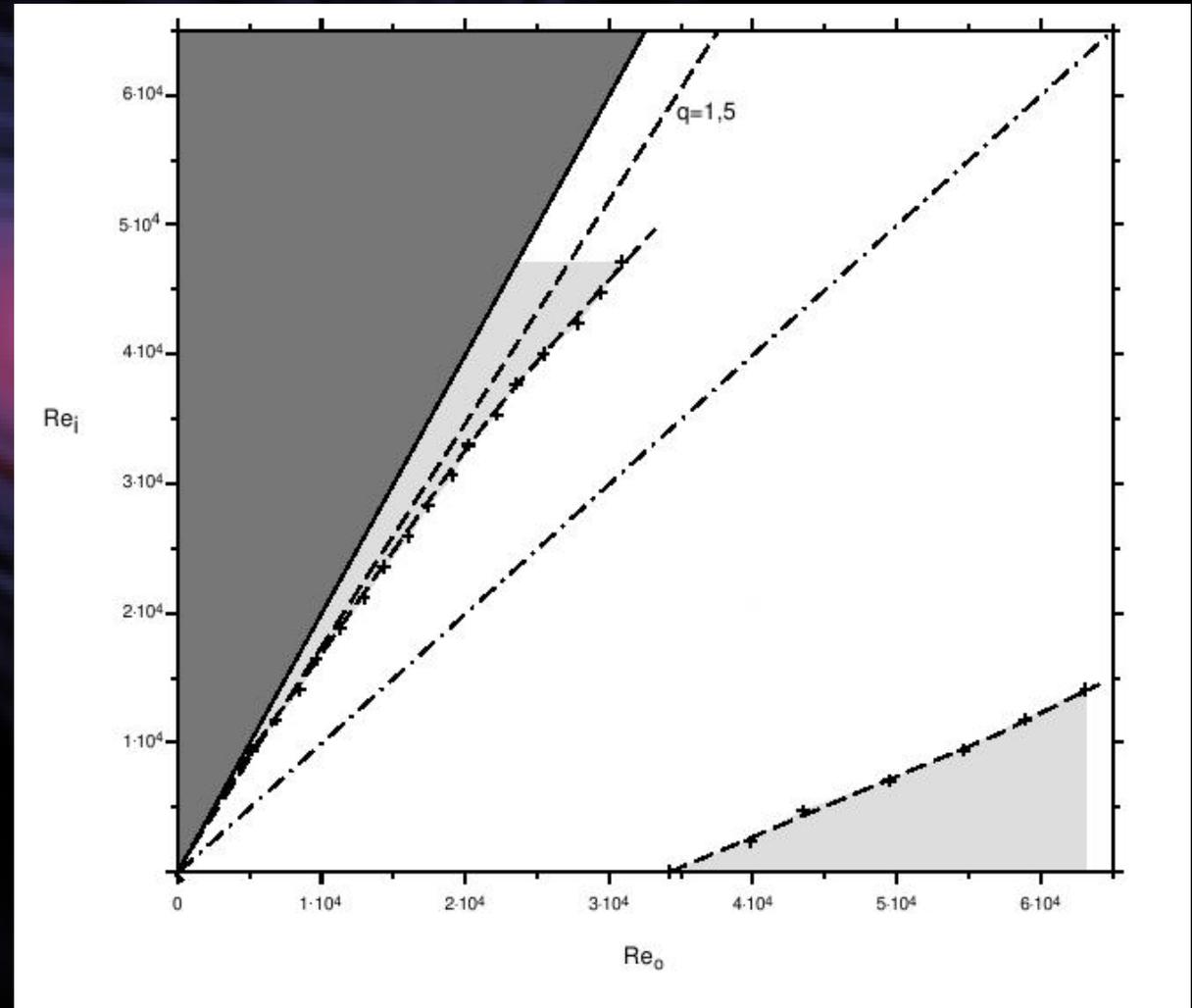

Conclusion : In a laboratory experiment (a not-so-close analog to a disk), differential rotation can give rise to turbulence despite published arguments.

**Differential Rotation may lead to turbulence in Keplerian disks.**

# Princeton Experiment
## (H.Ji, M.Burin, E.Schartman & J.Goodman, 2006)
### (Comments requested by meeting organizers...)

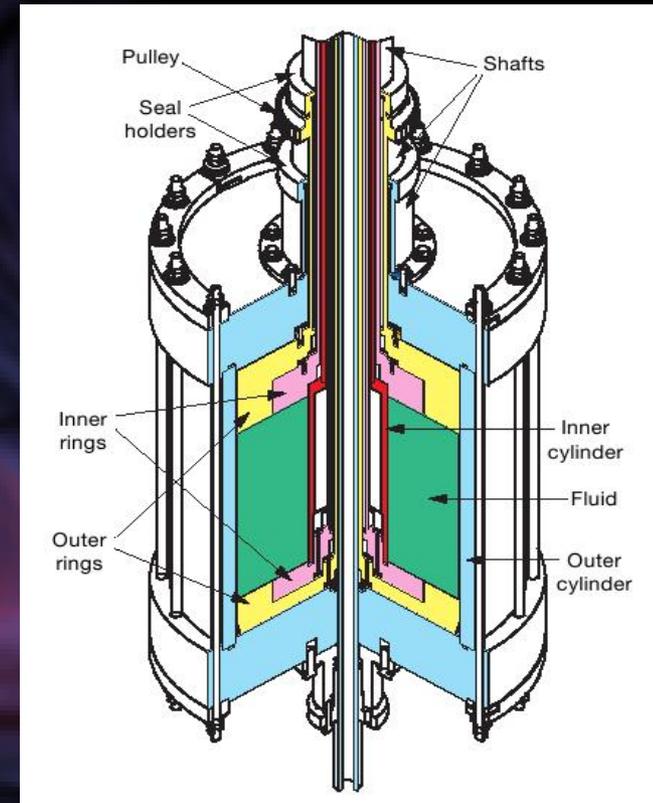

* **Couette-Taylor flow.**

* **No flow-visualization**, due to technical difficulties (private communication, M.Burin). Turbulent flows can not be positively differenciated from laminar flows. (As stated in the paper itself.) This is why Ji et al never claim that the flows are stable. They merely discuss fluctuation levels.

* **Boundary conditions :**
    * Stubby aspect ratio with independently rotating rings to compensate and avoid large scale circulation.

    * clever setup but :
        - experimental calibration does not agree with numerical simulations of this setup (see Burin et al, 2006).

        - calibration done "blind" (no visualization) and only close by the Rayleigh boundary where the flow could be turbulent (priv. comm. M.Burin). Should have been done in a regime where there are no doubts that the flow is laminar. Could have been returning the flow to a laminar regime after a sub-critical transition.

# Princeton Experiment
(H.Ji, M.Burin, E.Schartman & J.Goodman, 2006)
(Comments requested by meeting organizers...)

* Main argument for **Astrophysical flows** :

  * measured $\beta < 6.2 \cdot 10^{-6}$

  * then compares
    $\alpha$ and $\beta$ parameters
    numerically, by deriving
    a formulation of the $\alpha$ viscosity for the laboratory setup. Concludes that $\alpha$ and $\beta$
    should have a similar value, thus that $\alpha \sim \beta < 6.2 \cdot 10^{-6}$ is too small ($\alpha \sim 10^{-3}$).

  > Lastly, it is useful to relate the above upper bound for $\beta$ to the more commonly used Shakura–Sunyaev $\alpha$ parameter[1]: $v_{turb} = \alpha \Omega h^2$. We replace this by $v_{turb} = \alpha \Omega (r_2 - r_1)^2$ since $r_2 - r_1$ is smaller than $h$ in our experiment, and we presume that the dominant turbulent eddies scale with the smallest dimension of the flow ($h \ll r$ in most disks). Then $\alpha = \beta q \bar{r}^2 / (r_2 - r_1)^2 \approx \beta q$ (see Fig. 3 legend for a definition of $q$),

* BUT : this formulation for the $\alpha$ viscosity has meaning only in a thin keplerian disk where $C_s = \Omega \cdot H$. It makes absolutely no sense for this experimental setup.

* What makes sense is to compare $\alpha$ and $\beta$ value in an Astrophysical context.
  For $\alpha \cdot \Omega \cdot H^2 = \beta \cdot \Omega \cdot R^2$, then $\beta / \alpha = (H/R)^2$, therefore for $\alpha \sim 10^{-3}$, and $0.001 < H < 0.1$,

  $\beta \sim 10^{-9} - 10^{-5}$, still provides adequate transport of angular momentum.

# Conclusion

* The issue of **differential rotation is still very much open**.

* The debate about the origin of turbulence in disks should be a search to **characterize the turbulent state** of disks and to **model transport properties**. Today, it sometimes seems to be more an effort to **eliminate non-fashionable instabilities**.

* Considering the complexity of disks, it would be **naive to think** that **one process only** participate in angular momentum process.

* Discouraging work on various instabilities is **damaging for disk understanding**. Not only angular momentum is transported. Chemistry, planet formation, etc. are affected by turbulence that may not ultimately be relevant for angular momentum transport.

* Should be **acknowledged** :

   - The **limitations** of our tools : numerical, analytical, and experimental. Being able to describe a process better/more easily does not make it more relevant or important.

   - The **lack** of observational **constraints** on **disks physics**.

# Tools

Hopefully... ...but quite possibly... ...while most often presented as :

### Analytical model

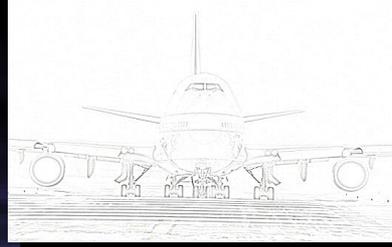 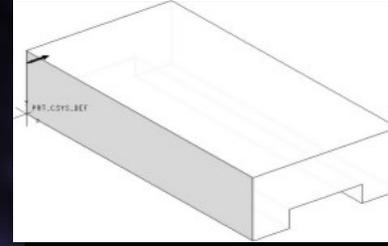 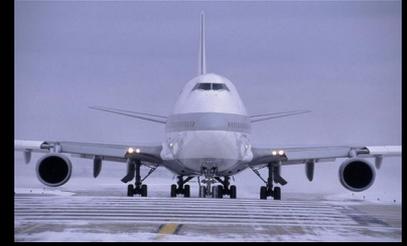

### Real object

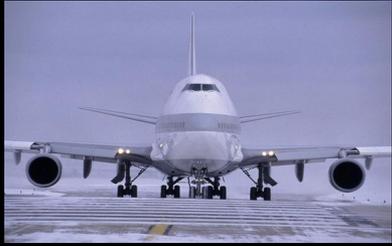

### LES / DNS models

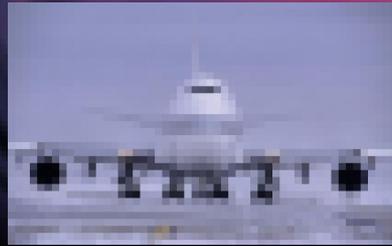 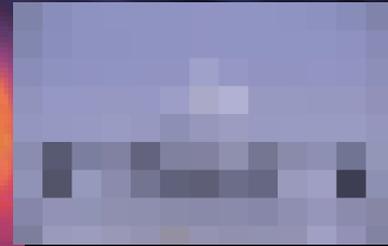 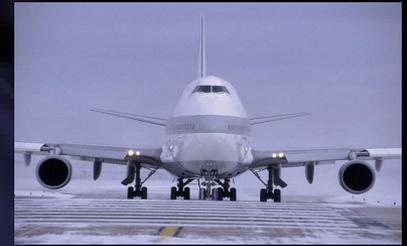

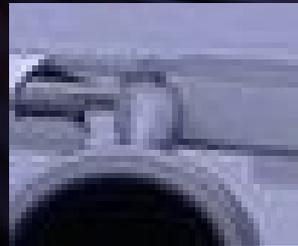 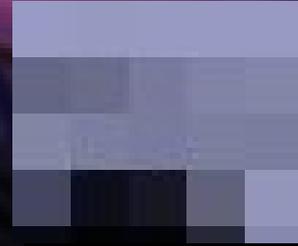 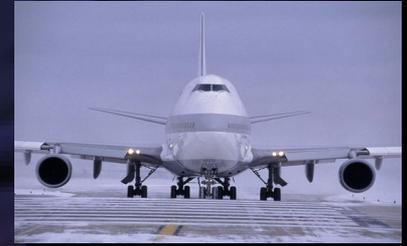

### Laboratory model

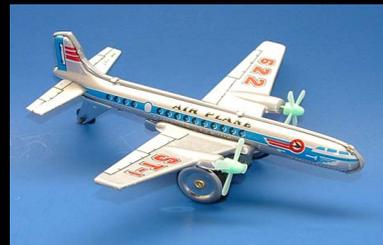 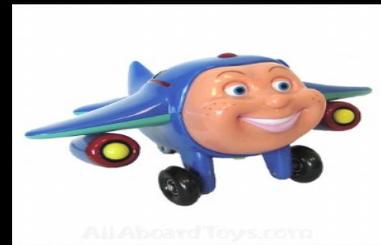 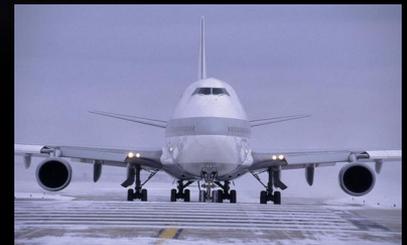